\documentclass[twocolumn, tighten]{aastex6}
\usepackage{amsmath}
\usepackage{amstext}
\usepackage{natbib}
\usepackage{graphicx}
\usepackage{acronym}
\usepackage{hyperref, url}

\usepackage[colorinlistoftodos]{todonotes}

\shorttitle{GALVIS}
\shortauthors{Tollerud \& Peek}

\newcommand{\HI}{\ion{H}{1}}
\newcommand{\MHI}{M_{\rm HI}}
\newcommand{\Mgas}{M_{\rm gas}}
\newcommand{\LCDM}{$\Lambda$CDM}
\newcommand{\kps}{{\rm km/s}}
\newcommand{\zre}{z_{\rm reionization}}
\newcommand{\Tvir}{T_{\rm vir}}

\newacro{GALFA-HI}{Galactic Arecibo L-band Feed Array \HI}
\newacro{ELVIS}{Exploring the Local Volume In Simulations}
\newacro{SDSS}{Sloan Digital Sky Survey}
\newacro{LG}{Local Group}
\newacro{LV}{Local Volume}
\newacro{MW}{Milky Way}
\newacro{HVC}{high-velocity cloud}
\newacro{CCC}{Compact Cloud Catalog}
\newacro{ALFA}{Arecibo L-Band Feed Array}
\newacro{PS1}{Pan-STARRS1}
\newacro{FAST}{Five Hundred Meter Spherical Telescope}
\newacro{LSST}{Large Synoptic Survey Telescope}

\begin{document}

%\title{Most Nearby Compact HI Clouds Are Not Local Dwarf Galaxies: A Signature of Reionization?}
\title{Where are All the Gas-Bearing Local Dwarf Galaxies? Quantifying Possible Impacts of Reionization}
%JEGP: Catchier?
\author{Erik J. Tollerud\altaffilmark{1}, J.E.G. Peek\altaffilmark{1, 2}}
\email{etollerud@stsci.edu}
\altaffiltext{1}{Space Telescope Science Institute, 3700 San Martin Dr., Baltimore, MD 21218, USA}
\altaffiltext{2}{Department of Physics \& Astronomy, Johns Hopkins University, 3400 N. Charles Street, Baltimore, MD 21218, USA}

\begin{abstract}
We present an approach for comparing the detections and non-detections of \ac{LG} dwarf galaxies in large \HI{} surveys to the predictions of a suite of n-body simulations of the \ac{LG}.  This approach depends primarily on a set of empirical scaling relations to connect the simulations to the observations, rather than making strong theoretical assumptions.  We then apply this methodology to the \ac{GALFA-HI} \ac{CCC}, and compare it to the \ac{ELVIS} suite of simulations. This approach reveals a strong tension between the na\"ive results of the model and the observations: while there are \emph{no} \ac{LG} dwarfs in the \ac{GALFA-HI}  \ac{CCC}, the simulations predict $\sim 10$.  Applying a simple model of reionization can resolve this tension by preventing low-mass halos from forming gas. However, and \emph{if} this effect operates as expected, the observations provide a constraint on the mass scale of dwarf galaxy that reionization impacts.  Combined with the observed properties of Leo T, the halo virial mass scale at which reionization impacts dwarf galaxy gas content is constrained to be $\sim 10^{8.5} M_\odot$, independent of any assumptions about star formation.

\end{abstract}

\maketitle

\acresetall{}

\section{Introduction}
\label{sec:intro}

Dwarf galaxies of the \ac{LG} provide uniquely faint limits, yielding a similarly unique range of constraints on galaxy formation and cosmology. These constraints typically hinge on connecting simulations in an assumed cosmology (usually the concordance \LCDM{} cosmology) to a particular model of galaxy formation, and comparing that in some manner to observations.  This general approach has led to a number of unexpected findings \citep[e.g.][]{klypinmsp, mooremsp, krav10satrev, BKBK11, brooks14}. However, these approaches have tended to focus on the stellar component of \ac{LG} dwarfs and their model equivalents. While this is a practical approach given that most dwarf galaxies of the \ac{LG} are passive and lack gas \citep{grcevich09, spekkens14}, the dwarf galaxy Leo T demonstrates that even ultra-faint dwarfs can have gas even to the present day \citep{ryanweber08}. Hence, it is important to consider what additional constraints may be gained by investigating the presence or absence of gas-bearing galaxies.

Recent advancements have improved the prospects for such investigations, especially in tracking the diffuse gas traced by \HI.  With the installation of the \ac{ALFA} at the William E. Gordon 305 meter antenna at the Arecibo Observatory, the rate at which Arecibo can survey the 21-cm line of \HI\ has increased nearly 7-fold. Arecibo provides the sensitivity of a single dish instrument unavailable to radio interferometers with an angular resolution ($\sim$ 1 kpc for objects 1 Mpc away) unavailable to smaller single-dish observatories. Therefore, the surveys conducted with ALFA provide the best available data sets for searching for these gas-rich dwarf galaxies in the local group. These surveys, tuned to compact \HI clouds, have found interesting populations of Galactic disk clouds mixed in with dwarf galaxy candidates.  Some of these have later been confirmed to be galaxies beyond the local group \citep{saul12, T15, Sand15}. This suggests that such \HI{} surveys may be useful for improving the understanding of nearby dwarf galaxies.

A major uncertainty to any investigation of the gas (or stellar) content of low-mass galaxies is the impact of reionization. Theory and simulations strongly suggest that dwarf galaxies with low enough mass dark matter halos are significantly affected by reionization, dramatically altering how the galaxies form \citep[e.g.][]{barkanaandloeb, okamoto08, ricotti09}.  This can have major observational consequences, such as explaining the relatively small number of dwarf galaxies in the \ac{LG} compared to the number of subhalos that are seen in \LCDM{} simulations \citep[e.g.][]{bullock00}.  While it is strongly suspected that these impacts are important, the details of these effects, particularly when and at what mass scale the effects become important, are not at all clear.  In this work we investigate whether simply positing the presence of such an effect combined with the empirically observed \HI{} content of \ac{LG} dwarf galaxies can provide independent constraints on the impacts of reionization on dwarf galaxies and their dark matter halos.

This paper is organized as follows:
In \S \ref{sec:galfa}, we describe the \ac{GALFA-HI} survey and the compact high-velocity \HI{} clouds it has identified as dwarf galaxy candidates;
In \S \ref{sec:dwarfgals} we describe optical searches for dwarf galaxies near the \HI{} clouds described in \S \ref{sec:galfa};
In \S \ref{sec:mockgalfa} we describe a method for creating mock sets of \ac{GALFA-HI} Compact Cloud Catalog galaxies, using the \ac{ELVIS} suite of simulations;
In \S \ref{sec:reionization} we describe how the combination of all of the above provides constraints on the scale at which reionization strips gas from dwarf galaxies;
and in \S \ref{sec:conc} we conclude.
To allow reproducibility, the analysis software used for this paper is available
at \url{https://github.com/eteq/galvis}.

\section{\ac{GALFA-HI} and its Compact Clouds}
\label{sec:galfa}

The \ac{GALFA-HI} survey comprises observations of neutral hydrogen with the 21-cm line taken with the 305 meter William E. Gordon telescope located in Arecibo, Puerto Rico. These observations began in 2004 with the installation of \ac{ALFA}, which provided an almost 7-fold increase in mapping speed over the L-band wide feed. \ac{GALFA-HI} observations have an angular resolution of 4$^\prime$, and a spectral resolution of 0.184 km s$^{-1}$, over the velocity range -650 to +650 km s$^{-1}$ V$_{LSR}$. This work relies on the observations compiled into the \ac{GALFA-HI} Data Release 1 (DR1): 3046 hours of observations covering 7520 deg$^2$ of sky between Declination -1$^\circ$ and 38$^\circ$ \citep{Peek11galfadr1}.

\citet{saul12} used \ac{GALFA-HI} DR1 to generate a catalog of 1964 compact \HI{} clouds using a template-matching technique, the \ac{GALFA-HI} \ac{CCC}. The search was designed to be sensitive to clouds smaller than 20$^\prime$ and with linewidths between 2.5 and 35 $\kps$.
The sensitivity of the search was measured empirically through a signal injection approach (section 3.3 of \citet{saul12}). Simulated clouds were injected with a range of positions, velocities, linewidths, sizes, aspect ratios, and brightnesses and the detection algorithm run. It was found that a rather simple function of these parameters was able to reliably predict whether the search algorithm could find a given cloud.

\citet{saul12} divided the 1964 clouds in the \ac{CCC} into a number of categories based on their linewidth, position, and velocity. Among these categories, we are interested in two in particular, which contain all 719 the of the clouds with $|v_{LSR}| > 90 \; \kps$. All clouds that are close in position velocity space to large \acp{HVC} known in the \citet{WvW91} catalog  are categorized as \acp{HVC}, while those far from these clouds are called ``Galaxy Candidates.''
This distinction was made under the assumption that small \HI{} clouds near larger high velocity clouds are much more likely to be \acp{HVC} than to be galaxies. Practically, this proximity is quantified using the parameter
\begin{equation}
D = \sqrt{\left(\Theta^2 + f^2\left(\delta v\right)^2 \right)},
\end{equation}
where $\Theta$ is the angular distance between two clouds, $\delta v$ is the LSR velocity difference, and $f$ is 0.5 degrees / km s$^{-1}$ \citep{Peek09}. For each cloud in \citet{saul12}, this parameter is measured against all \acp{HVC} in the \citet{WvW91} catalog, and those clouds with minimum D $>$ 25$^\circ$ are classified as Galaxy Candidates. This procedure finds a total of 27 such candidates.

This method of distinguishing between likely HVCs and possible dwarf galaxies is supported by work by \citet{meyer15}, who showed that including  D $<$ 25$^\circ$ clouds  diluted an \HI-based search for more distant UV-bright galaxies.
Our goal in this work is to identify which of these candidates are actually \ac{LG} dwarf galaxies (\S \ref{sec:dwarfgals}), and create a mock version of this survey using simulations in an appropriate cosmological context (\S \ref{sec:mockgalfa}).

\section{Dwarf Galaxies in \ac{GALFA-HI}}
\label{sec:dwarfgals}

The set of Galaxy Candidates we describe in \S \ref{sec:galfa} require follow-up to determine which, if any, have optical galaxies within the \ac{GALFA-HI} beam.
It is precisely this that was the goal of the surveys described in \citet{T15} and \citet{Sand15}.
These samples, as described in more detail in Bennet et al. (in prep), observed the \ac{GALFA-HI}
 \ac{CCC} Galaxy Candidate fields and resolve stars to a depth of $\sim 2$ mags deeper than \ac{SDSS}.
 These surveys determined that 5 have optical counterparts that are likely nearby galaxies, but they are at $2$ to $10$ Mpc \citep{T16}, placing them in the \ac{LV} rather than the \ac{LG}.

While the description above focuses on new unknown galaxies, the \ac{GALFA-HI} footprint also contains nearby dwarf galaxies that are not in the \ac{CCC} simply because they were previously-known.  Specifically, Sextans B, GR 8, and KKH 86 are in the \ac{GALFA-HI} footprint.
However, as with the confirmed galaxy candidates described above, and as discussed in more detail in \citet[][Figure 5]{mcconlgcat}, all three of these galaxies are clearly in the Hubble Flow (both by distance and velocity), and therefore not a part of the \ac{LG}.

Hence, while there are a large set of \ac{GALFA-HI} \ac{CCC} Galaxy Candidates, \emph{no} \HI{}-bearing \ac{LG} galaxies exist in the \ac{GALFA-HI} footprint. This then raises the question of whether or not such galaxies might be \emph{expected} by galaxy formation in an appropriate cosmological context.  It is to this question that we turn in the following sections.

\section{Constructing a Mock \ac{GALFA-HI} Survey}
\label{sec:mockgalfa}

We begin by asking how many dwarf galaxies we would expect in \ac{GALFA-HI} given a simple set of galaxy formation and cosmological assumptions (i.e., \LCDM{}).   We start with the assumption that all dwarf galaxies are contained inside of dark matter halos \citep[e.g.][]{willmanstrader12}. This allows our starting point to be \LCDM{} halo catalogs in an environment comparable to the \ac{LG}.

The specific set of halo catalogs we use for this experiment are taken from the \ac{ELVIS} suite \citep{gk14elvis}.  These simulations were designed to simulate environments similar to the \ac{LG} in the sense of having two $M_{\rm halo} \sim 10^{12}$ halos at distances and relative velocities similar to the \ac{MW} and M31.  We use this suite of simulations, because accounting for \emph{both} $\sim M_*$ galaxies of the \ac{LG} is critical: a significant part of the \ac{GALFA-HI} footprint is in the direction of the M31 halo on the sky.  \citet{gk14elvis} demonstrated that the non-satellite dwarf galaxy halos in \ac{LG}-like environments are significantly different than individual isolated $\sim M_*$ galaxy halos.  Hence, including both galaxies in their correct orientations relative to the simulated \ac{GALFA-HI} footprint is important for generating a realistic estimate.  The \ac{ELVIS} suite provides just this, with 12 \ac{LG}-like analogs (i.e., 24 total $\sim M_*$ galaxies and their attendant dwarf halos).

The \ac{ELVIS} suite comprises dark matter-only simulations.  \ac{GALFA-HI} is only sensitive to $\MHI$ (and our optical observations detected $M_*$), so we must impose some model on the simulations to determine the observables for a given dark matter halo in \ac{ELVIS}.  Such models can span a wide range of complexity and assumptions, from full cosmological hydrodynamic simulations to basic semi-analytic approaches \citep[e.g.][]{stewart09, galacticus, rod12, sawala14, snyder15, wheeler15}.  Here we consider a simple, primarily empirical model.  While such a model is unlikely to capture the complex physics of galaxy formation in detail, our goal is a rough comparison with the \ac{GALFA-HI} observables; a detailed investigation of galaxy formation models is beyond the scope of this work.

\begin{figure}[htb]
\begin{center}
\includegraphics[width=1\columnwidth]{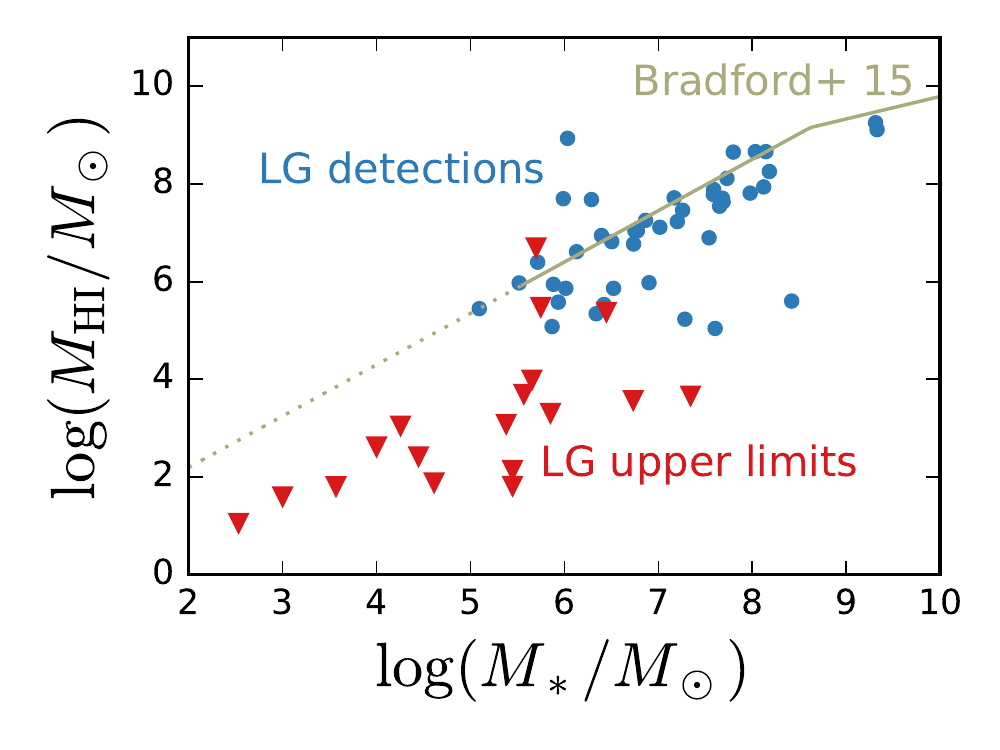}
\caption{$M_*$-to-$\MHI$ relation of \citet[][black lines]{bradford15}, along with \ac{LG} dwarf galaxies. The solid line shows the approximate region where the \citet{bradford15} relation is directly calibrated, while the dotted section is extrapolation.  The (blue) circles show non-satellite \ac{LG} dwarfs with \HI{} detections from the \citet{mcconlgcat} compilation.  The (red) triangles show \emph{upper limits} on \HI{} from \citet{spekkens14}, which are are all satellite galaxies.  This demonstrates that, while not directly calibrated in this region, the \citet{bradford15} relation is moderately consistent with the \ac{LG} galaxies with \HI{}, but lies well above the limits for \ac{LG} satellites.}
\label{fig:MstarMHI}
\end{center}
\end{figure}

Our model is as follows: we begin with positions, velocities, and halo masses provided in the public data release of \ac{ELVIS}\footnote{\url{http://localgroup.ps.uci.edu/elvis/}}.
To obtain stellar masses for each halo, we apply the abundance-matching based halo-to-stellar mass relation of \citet{gk14elvis} to obtain $M_*$ for each halo.  This has the specific advantage of being calibrated to both the \ac{LG} and \ac{ELVIS}, precisely the two data sets we are interested in here.  For considerations of completeness of our optical follow-up, we convert to luminosity assuming a mass-to-light ratio of unity and a standard $r-band$ bolometric correction \footnote{\url{http://mips.as.arizona.edu/~cnaw/sun.html}}. To determine $\MHI$ we then use the $M_*$ in combination with the $M_*$-to-$\Mgas$ relation of \citet{bradford15}.  We convert this relation from $\Mgas$ to $\MHI$ by inverting the procedure described in \citet{bradford15}. This procedure is straightforward as the observations from that work are also \HI{} observations. In Figure \ref{fig:MstarMHI} we show this $M_*$-to-$\MHI$ relation (black lines), along with \ac{LG} dwarf galaxies with \HI{} from \citet{mcconlgcat} and limits from \citet{spekkens14}. This figure demonstrates that, even beyond where it is calibrated, the \citet{bradford15} is consistent with the \ac{LG} dwarfs (although with scatter comparable to the \citealt{bradford15} dataset). While there may be a small bias in the relation relative to the \ac{LG}, the \citet{mcconlgcat} compilation is inhomogeneous enough that we opt to stick with the \citet{bradford15} extrapolation with no corrections, as it is a much more homogenous dataset.  We also see, from the \citet{spekkens14} dataset, that \emph{satellites} in the \ac{LG} are clearly \HI{} deficient relative to the \citet{bradford15} relation. This fact motivates our choice to remove satellites from the mock \ac{GALFA-HI} \ac{CCC} galaxies, described in more detail below.

The above procedure results in a set of 12 halo catalogs of \ac{LG} analogs, with halo, stellar, and \HI{} masses for each. We now describe how we convert these catalogs into mock \ac{CCC} galaxies that would be found in mock \ac{GALFA-HI} surveys.

The first step is identification of the massive halos in the catalog with the \ac{MW} and M31. While the \ac{LG} total mass is relatively well constrained and the \ac{MW} and M31 are clearly dominant, exactly how to apportion the mass between the two is relatively uncertain \citep[e.g.,][]{klypin02, watkins10, toll12, gonzalez14, pen14}.  Hence, to marginalize over this uncertainty and at the same time boost our count of mock surveys, we create a mock \ac{GALFA-HI} footprint \emph{twice} for each \ac{ELVIS} pair, swapping which large halo hosts the \ac{MW} and M31.  While this does reuse some of the same halos twice, the details of detectability and different distances from the two hosts means that each mock survey is a  different sample, and there is therefore a relatively weak covariance between the two mock surveys from the same \ac{ELVIS} pair. While the presence of weak covariance means these are not 24 completely independent samples, the weakness of the covariance means there is substantially more power in using all of the halos individually instead of only one of each pair.  Hence, the procedure described below is repeated for each of the 24 \ac{MW}/M31 pairs to produce our ensemble of mock surveys.

For each pair, we fix the orientation of the halos on the mock sky by placing the mock Earth in the unique position and orientation where the distance to the center of the \ac{MW} halo is 8.5 kpc (IAU standard), the center of the \ac{MW} halo is in the direction of the origin, and the center of the M31 halo is in the direction of $l=121.17^{\circ}$, $b=-21.57^{\circ}$ (M31 in Galactic coordinates).  In that orientation, we add a $220 \kps $ velocity offset to each halo in the $l=90^{\circ}$ direction to model the Sun's motion around the Galactic center (IAU standard).  This yields a mock survey footprint in Galactic coordinates with radial velocities and distances as would be observed from the real Earth.

To determine detectability of a galaxy in the \citet{saul12} \ac{CCC} from the mock survey, we overlay the footprint (and spatially variable depth) of the real \ac{GALFA-HI}, and identify the nearest pixel to each halo.  Using the distance for that halo and its $\MHI$ from above, we compute its expected \HI flux, and we accept it if it is higher than the detectability threshold of the \ac{CCC} (described in \S \ref{sec:galfa}).  We further apply a $|v_{LSR}| > 90 \; \kps$  cut to match the galaxy candidate sample described in \S \ref{sec:galfa}.

We also consider a final cut on \emph{optical} detectability in the follow-up observations by assuming $M_*/L_r \sim 1$, and cutting on the $r$-band detection thresholds for follow-up observations.
However, this cut is less stringent than the $\MHI$ sensitivity cut based on our thresholds from \S \ref{sec:dwarfgals}, and hence has no effect on the final count as described below.
We leave this cut in as a parameter in the model, however, and investigate its effects further in \S \ref{sec:reionization}.

This yields our sample of mock dwarf galaxies that could be found in the \ac{CCC} under the assumption that \emph{all} galaxies have their full allotment of \HI. However, Figure \ref{fig:MstarMHI} demonstrates clearly that this is not the case: in the \ac{LG}, most lower-mass \emph{satellite} galaxies have orders-of-magnitude less \HI\ than star-forming galaxy scaling relations imply.  The exact process by which theses satellites are quenched is not certain.
Whether non-satellite dwarfs self-quench is an open question, but at least in the \ac{LG} it seems likely to be quite rare given that nearly all dwarfs beyond the \ac{MW} and M31 have gas \citep[e.g.,][]{mcconlgcat, wetzel15}.
But it is clear that some mechanism removes gas and therefore quenches star formation in \emph{satellites} \citep[e.g.][and references therein]{spekkens14, wetzel15, fill15, simpson17}.

\begin{figure*}[htb]
\begin{center}
\includegraphics[width=1.4\columnwidth]{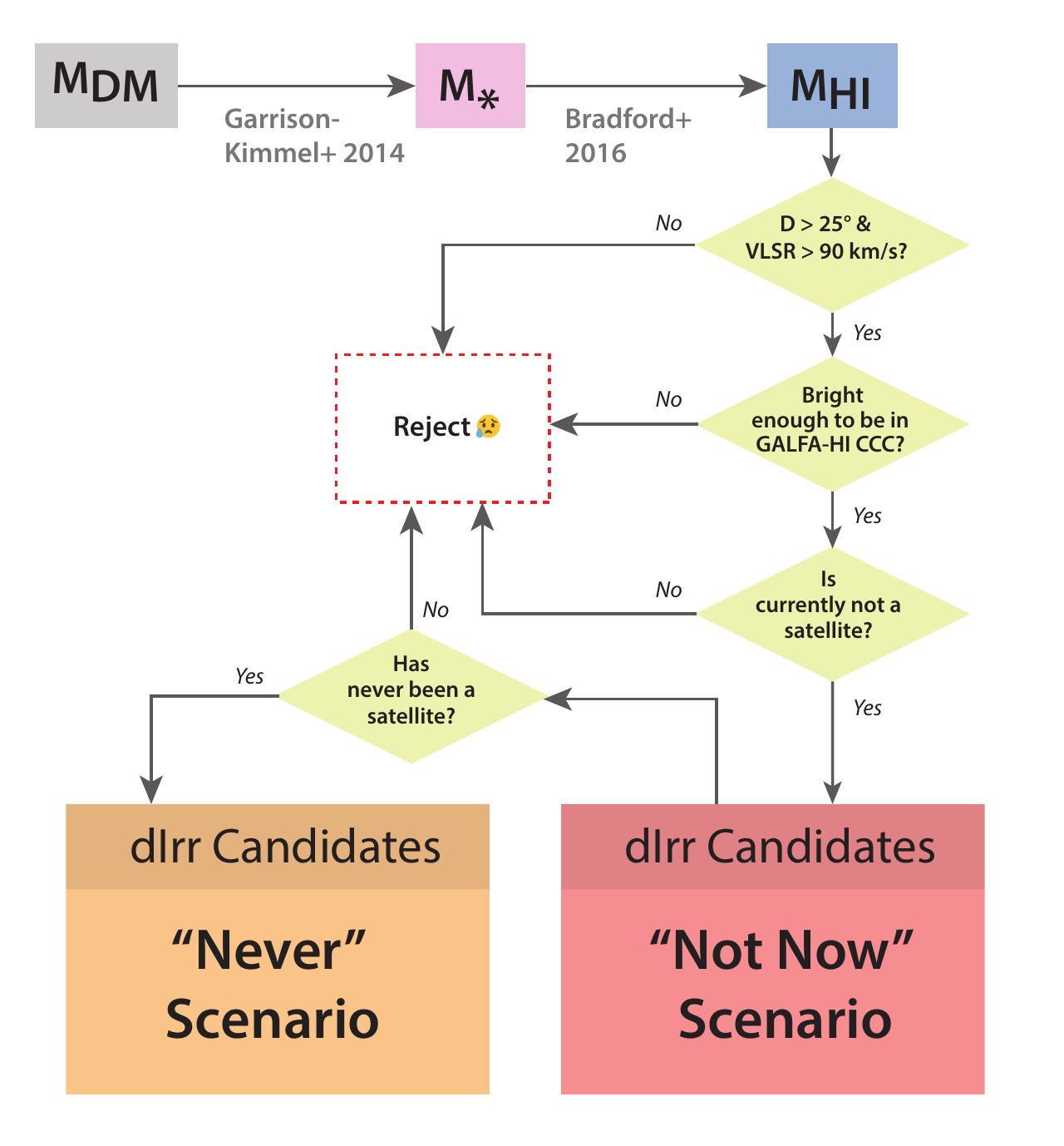}
\caption{Schematic diagram of the method to construct the dwarf galaxy candidate samples (described in detail in \S \ref{sec:mockgalfa}).}
\label{fig:elvistogalfacartoon}
\end{center}
\end{figure*}

Here, we consider two simple scenarios intended to bracket various host-driven quenching mechanisms and timescales. The two scenarios, as well as the approaches described in this section as a whole, are illustrated in Figure \ref{fig:elvistogalfacartoon}. In Scenario ``Not-Now'', we assume any galaxy that is a subhalo\footnote{Subhalos are defined here following the \ac{ELVIS} catalogs, based on the 6D friends-of-friends Rockstar halo finder \citep{rockstar}.} of the \ac{MW} or M31 at $z=0$ has lost its gas, and all others are normal. In Scenario ``Never'', we assert that galaxies immediately lose all their \HI\ gas the moment they become subhalos at \emph{any} time.  Note that this makes Scenario ``Never'' a strict subset of ``Not-Now''.  While neither of these scenarios are likely to be correct in detail, and do not allow for self-quenching, they bracket many scenarios based on direct physical effects between a host and its subhalos, and hence serve for our current purpose of providing an estimate of what we would expect \ac{GALFA-HI} to see.

\begin{figure*}[htb]
\begin{center}
\includegraphics[width=1\textwidth]{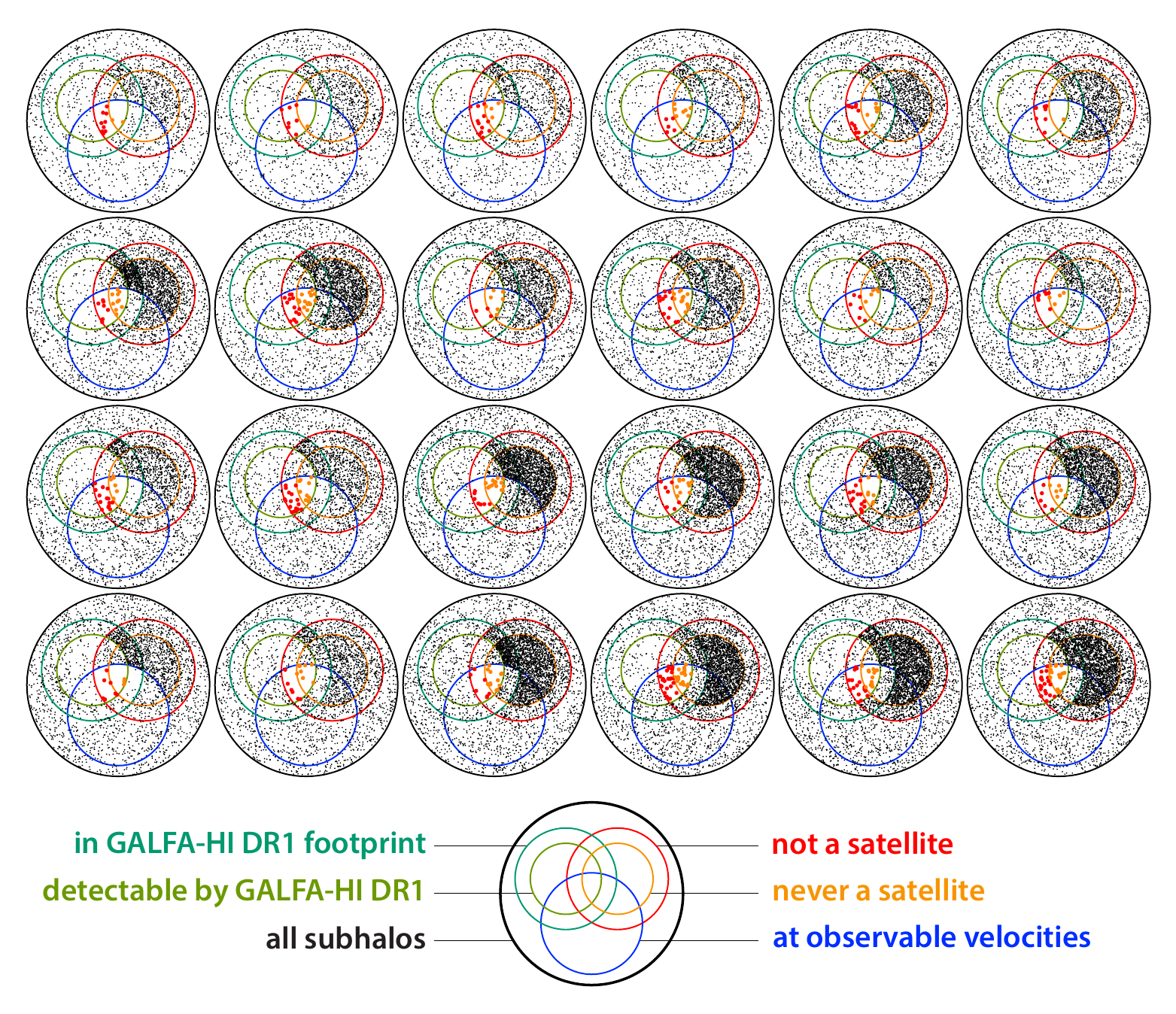}
\caption{Number of halos in each of the \ac{ELVIS} hosts for the sample cuts described in the text. The number of points in each circle/overlap region demonstrates the effect of each cut in limiting the sample of candidate galaxies.  The red and orange (larger) points correspond to halos for scenario ``Not-Now'' and ``Never'', respectively. }
\label{fig:elvismallmultiples}
\end{center}
\end{figure*}

In Figure \ref{fig:elvismallmultiples} we show the counts of halos in each of the 24 mock \acp{CCC} (i.e., each of the \ac{ELVIS} host halos as the MW).  Each point represents an individual halo, and points inside a given circle are those that pass the corresponding cut.  One clear point this Figure demonstrates is the importance of the velocity and detectability cuts -- together they remove a $\gtrsim 50$\% of the sample, underscoring the importance of correctly modeling the specific  observational details of \ac{GALFA-HI} and the \ac{CCC} detection algorithm. Also important here is the recognition that there is quite a lot of variability in these samples, both in the total number of halos and the effects of the cuts.  This represents a mix of true cosmic variance as well as uncertainty in the \ac{MW} and M31's properties, encoded in the spread of properties for the \ac{ELVIS} host halos.

\begin{figure*}[htb]
\begin{center}
\includegraphics[width=0.8\textwidth]{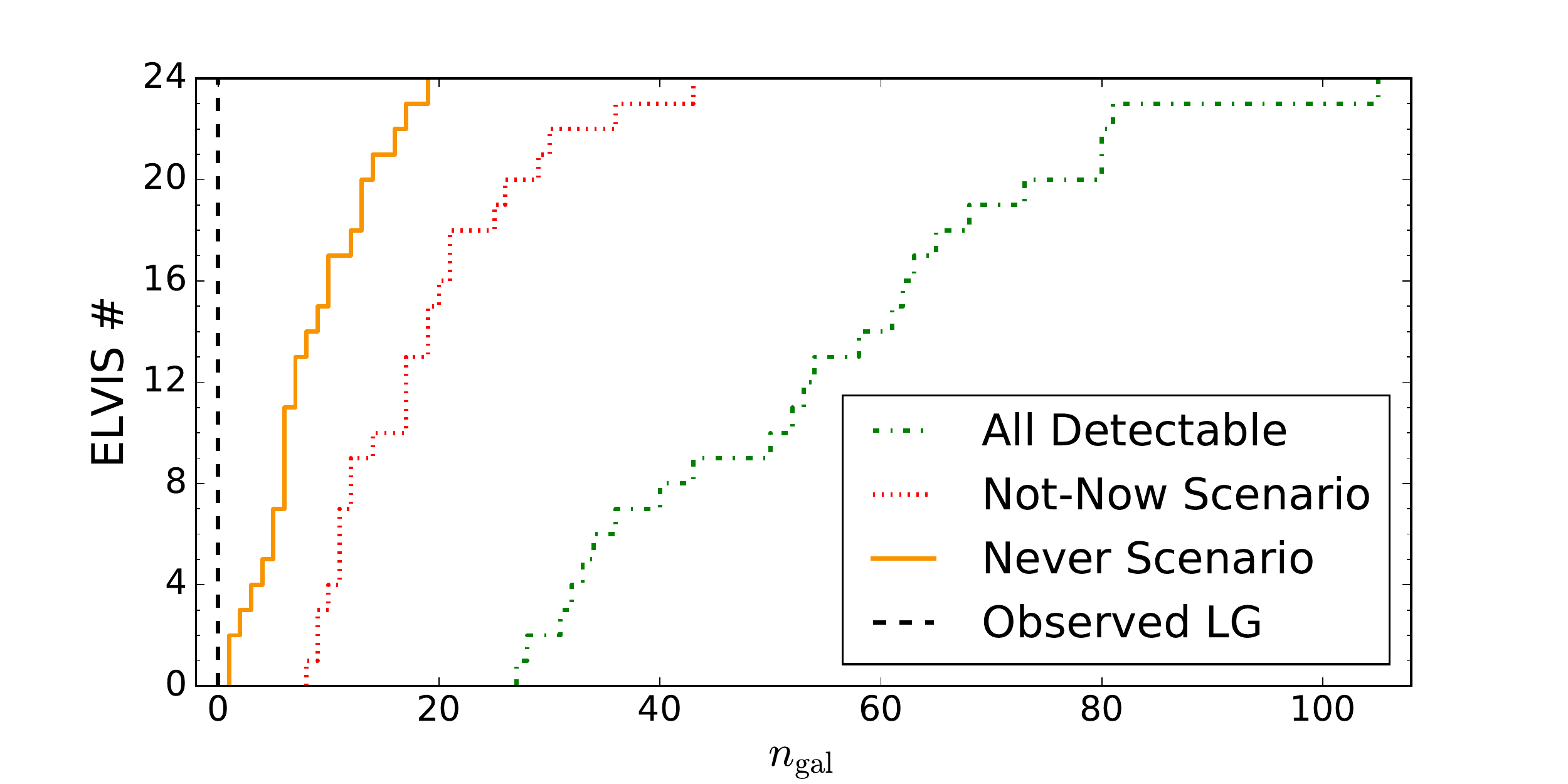}
\caption{Summary of the expected dwarf irregular candidates predicted following the mock \ac{GALFA-HI} \ac{CCC} procedure outlined in \S \ref{sec:mockgalfa}.  Lines show the cumulative distribution of the number of galaxies for the ``Never'' Scenario (orange solid), ``Not-Now'' Scenario (red dotted), and with no cut on satellites (green dot-dashed).  Also shown is the observed number of \ac{LG} galaxies in the {\emph real} \ac{GALFA-HI} \ac{CCC} as the dashed (black) vertical line: zero. This demonstrates the stark difference between the observations and simulation predictions for the scenarios described here.
}
\label{fig:countsummary}
\end{center}
\end{figure*}

In Figure \ref{fig:countsummary}, we summarize the cumulative distribution functions for predicted dwarfs in the footprint over the 24 mock surveys.  Unsurprisingly, for the scenario where satellites are included (green dot-dashed), there are many satellites predicted.  However, as discussed above, this case is already ruled out by the observation that most \ac{LG} satellites lack observable \HI{}.
More striking are the lines for the Not-Now (red dashed) and Never (solid orange) scenarios.  While the numbers are much reduced relative to the All scenario, the typical number counts are in the $5-20$ range.  In contrast, the observations (discussed in Section \ref{sec:dwarfgals}) show \emph{zero} galaxies. \emph{In fact, none of the mock surveys in the scenarios outlined above are consistent with the observations.}

The process outlined in this section demonstrates that, by constructing a mock \ac{GALFA-HI} catalog from the \ac{ELVIS} simulations, we determine that the \ac{GALFA-HI} observation of no \ac{LG} dwarfs is quite surprising at face value.
Several caveats apply, however.
First, because there are only 12 realizations of \ac{ELVIS} simulations, we only have 24 mock catalogs.
Hence, it is possible that the \ac{LG} is simply a $<$ 1 in 24 ($\sim 2 \sigma$) outlier.
This cannot be ruled out without more \ac{ELVIS}-like simulations, but is at least suggested against by the magnitude of the discrepancy outlined above.
Second, this could be interpreted as evidence that \LCDM{} (the underlying cosmology for \ac{ELVIS}) is not an accurate description of small-scale structure, like some interpretations of the ``missing satellites problem'' \citep{gotzwdm, rochasidm}, but here specific to gas-bearing dwarfs. Third, because \ac{ELVIS} is a collisionless dark matter simulation, it does not contain any baryons effects.
Hence, baryonic or hydrodynamic effects that might suppress the numbers of dwarf galaxies are not accounted for \citep[e.g.][]{pontzen12, governato12, brooks13}.
Most work on this topic has focused on how this impacts the \ac{LG} satellites, however, which we explicitly excise from the sample.
It is unclear which, if any, of these mechanisms apply for non-satellites like those considered here, and investigating such effects is beyond the scope of this paper.
Finally, it is possible that the $\Mgas$-to-$M_*$ relation of \citet{bradford15} does \emph{not} extend to below $M_*<10^6 \, M_{\odot}$ (i.e., the extrapolation in Figure \ref{fig:MstarMHI}).  If there is a break in this relation, the number of dwarfs with gas would be suppressed, solving the aforementioned tension.  It is precisely this possibility that is described in the next section, taking the causative mechanism to be reionization.

\section{Implications for Reionization}
\label{sec:reionization}

To estimate the effects of reionization we now consider a minimal toy model of the effect of reionization on dwarf galaxies, inspired in part by the approach of \citet{bk14}.
Of course, there are a wide range of models for reionization and its effect on dwarf galaxies, far more thorough than that used here \citep[e.g.,][]{barkanaandloeb, gnedin00, okamoto08, bovill11, fitts17}.
We use the model described here primarily because it is both simple and can provide a direct probabilistic mapping between \ac{ELVIS} and the \ac{GALFA-HI} observations.

\subsection{Lower Limits from \ac{GALFA-HI} and \ac{ELVIS}}
\label{ssec:galfalimit}

Our model assumes there exists a characteristic halo virial mass ($M_c$) at a particular redshift ($\zre$).  Halos with a virial mass below $M_c$ at $\zre$ have their gas entirely removed (by $z=0$), and those above have their gas content unaffected by reionization.
 While such a sharp break in the $\MHI$-to-$M_{\rm halo}$ relation is unphysical, the subsequent evolution in $M_{\rm halo}$ from $\zre$ to $z=0$ has the effect of smearing the break over $\sim 1$ dex in $M_*$ at $z=0$, where our comparison to observations is performed (see \S \ref{ssec:leotlimit}). This model is implemented in the context of the formalism of \S \ref{sec:mockgalfa} by identifying the main-branch progenitors of the $z=0$ halos at a particular $\zre$. Those with a virial mass below $M_c$ are flagged as undetectable in \ac{GALFA-HI} due to the removal of their gas.

This model is flexible enough to immediately solve the problem posted at the end of \S \ref{sec:mockgalfa}: if $M_c$ is high enough, no \ac{LG} dwarf galaxies will have gas at $z=0$, thereby solving the apparent tension between the \ac{ELVIS} model and the observations.
With that in mind, we now turn to asking the probabilistic question of what $M_c$ is, given the observation that there are no \ac{LG} dwarfs in the \ac{GALFA-HI} \ac{CCC}.
To obtain this probability distribution, we start from the process of \S \ref{sec:mockgalfa} applied to each of the \ac{ELVIS} hosts for both the ``Never'' and ``Not-Now'' scenarios, for a range of optical $r$-band follow-up detection limits.  We apply the reionization model described above over a grid of $M_c$ and $\zre$ values of  $z\sim6.3, 7.1, 8.1,$ {\rm and} $9.3$ (set by available \ac{ELVIS} timesteps).
We then ask what fraction of the 24 hosts yield galaxy number counts consistent with the observations (zero), and consider this to be proportional to the probability density $P(M_c, r_{\rm lim}, \zre | 0)$.
We then marginalize the probabilities over the available $\zre$ values to provide estimates of $M_c$ (and the $r$ limit).

To do the marginalization we assume a (discrete) uniform distribution of the $\zre$'s available, because our goal is to provide estimates relatively agnostic to assumptions about $\zre$.
However, a more specific reionization model would likely provide a more peaked $\zre$ distribution, and therefore provide tighter constraints than we obtain here.
Relatedly, we note in passing that a different choice of marginalization (and stronger assumptions) could yield a different inference: marginalizing over over $M_c$ to instead estimate $\zre$.
For the purposes of this paper, we opt not to do this because we are more interested in $M_c$, and our $\zre$ grid is quite coarse, but this methodology could be straightforwardly applied to a more specific galaxy formation model's $M_c$ prediction, yielding a probability distribution over $\zre$.

\begin{figure}[htb]
\begin{center}
\includegraphics[width=1\columnwidth]{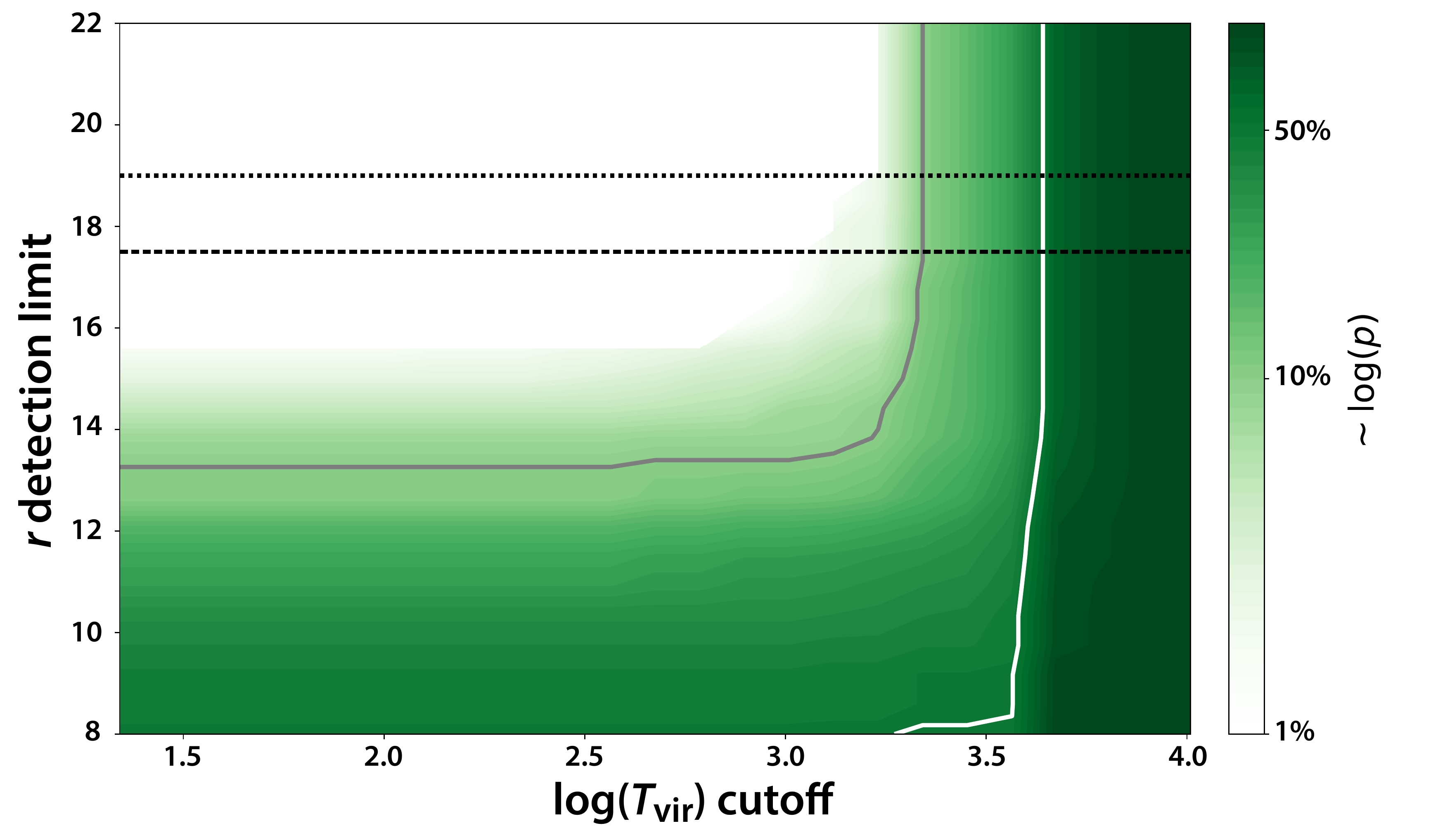}
\caption{Joint probability density for the halo mass at which reionization remove all gas and  optical detection limit in the follow-up observations.  The characteristic mass $M_c$ is converted to virial temperature $\Tvir$ in this plot for comparison to other literature.  The colored regions are proportional to the probability density, and the gray and white lines give the $10\%$ and $50\%$ contours. The horizontal dotted (lighter) line is the estimated follow-up detection limit described in \S \ref{sec:dwarfgals}. The  dashed (darker) line is a more conservative detection limits based on actual detected galaxies from \citep{T15}. This demonstrates both that the $M_c$ distribution has little covariance with the follow-up detection limits, and that the follow-up limits are deep enough that optical non-detections are unlikely to impact the inferred $M_c$.}
\label{fig:reionization}
\end{center}
\end{figure}

Figure \ref{fig:reionization} shows the result of the above procedure for the ``Not-Now'' scenario.
The $M_c$ values are converted to virial temperatures in this figure for comparison with literature values that are often reported as $\Tvir$.
It is immediately apparent that the probability density goes to $\sim 1$ at high $\Tvir$/$M_c$ values.
This is a result of the effect described above, that an arbitrarily large $M_c$ will \emph{always} be consistent with the observational result of no dwarf galaxies, as all of the candidates are removed by reionization.
Figure \ref{fig:reionization} also shows horizontal lines for two optical detection thresholds: the upper one corresponds to our estimated follow-up limits (\S \ref{sec:dwarfgals}), and the lower one is a highly conservative estimate based on the actual detected non-\ac{LG} dwarfs from \citet{T15} and \citet{Sand15}.
The corresponding probabilities for $\Tvir$ along those limits are essentially identical, showing that even if our follow-up detection limits are overly-optimistic, the key results of this work still hold.
Figure \ref{fig:reionization} also shows that the cutoff $\Tvir$ has a $\sim 50\%$ of being at least $10^{3.6} \; M_\odot$, or $M_c \gtrsim 10^{9} \; M_\odot$, consistent with simulation predictions of $M_c$ \citep[e.g.,][]{okamoto08}.

\subsection{Upper Limits from Leo T}
\label{ssec:leotlimit}

The procedure above provides only a \emph{lower limit} on $M_c$, because the \ac{GALFA-HI} observation here is the lack of \emph{any} detected \HI{}-bearing galaxies. From these observations alone, $M_c$ could be arbitrarily high, as this would still yield no observable \HI{}-bearing galaxies.  But of course, observations of the wider universe, indeed even other observations in the \ac{LG}, provide evidence of the existence of galaxies with \HI.  Applying the Copernican principle, it is reasonable to then use the existence of such galaxies in the \ac{LG} as a \emph{joint} constraint with the \ac{GALFA-HI} observation to achieve an actual estimate of  $M_c$.  Leo T provides the ideal constraint: it is both a dwarf galaxy in the \ac{LG} (at $\sim 400$ kpc from the \ac{MW}, not a satellite as we have defined it here) and the lowest-mass known \HI{}-bearing galaxy \citep{ryanweber08, weisz12}. It therefore provides the best source for an \emph{upper limit} on $M_c$.  Combining this constraint with the \ac{GALFA-HI} observations can then provide an estimate of $M_c$ (rather than just a limit).

Creating an upper limit on $M_c$ based on the existence of Leo T requires an estimate for the virial mass of the halo hosting Leo T at the time of reionization. To make such an estimate we start from the present day luminosity of Leo T from \citet{dejong08}, converted to a stellar mass ($1.4 \times 10^5 \; M_\odot$). We then find all the $z=0$ halos from the model outlined in \S \ref{sec:mockgalfa} that have stellar masses within $\pm 20 \%$\footnote{This specific percentage was chosen as the needed minimum to include enough halos to adequately sample the  probability distribution. A slightly wider or narrow range in stellar mass showed no signs of systematic bias in the center or width of the distribution.} of our Leo T estimate. For those halos we then identified the main progenitor at $\zre$ and adopt that as a possible $M_c$ limit.  This procedure provides an estimate of the scatter in the possible virial mass at $\zre$ of  Leo T due \emph{only} to uncertainty in its merger history. Other sources of scatter may contribute to the effects of reionization on Leo T-mass galaxies, possibly quite significantly \citep[e.g.][]{fitts17}.  However, we find that relatively large changes in the stellar mass assumed here yield negligible changes to the width of the distribution of $\zre$ halo masses.  This implies that the primary impact on the $\zre$ virial mass of Leo T \emph{itself} is driven by uncertainty in mapping any individual present day halo back to $\zre$.  We therefore adopt this merger-history driven scatter as the sole source of scatter for the purposes of this $M_c$ estimate.

\subsection{Combined Limits}
\label{ssec:combinedlimit}

\begin{figure}[htb]
\begin{center}
\includegraphics[width=1\columnwidth]{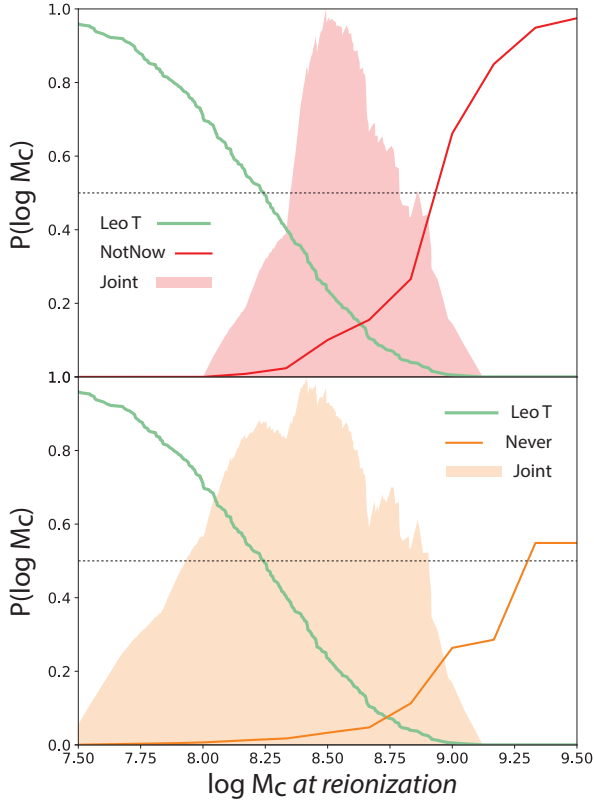}
\caption{Probability distributions of the critical mass for retaining \HI{} ($M_c$) at the time of reionization (which is a marginalized parameter in this model but $\sim 6$). The two \emph{limits} are shown as solid lines (their derivations are discussed in \S \ref{ssec:galfalimit} and \S \ref{ssec:leotlimit}). The upper panel is for the ``Not-now'' scenario when inferring the \ac{GALFA-HI} limits, and the lower panel is for the ``Never'' scenario.  Note that the limits as shown are $P(<\log{M_c})$ or $P(>\log{M_c})$, which is why the probabilities asymptote to 1.  The filled distribution shows the \emph{joint} constraint set by the two limits (discussed in more detail in \S \ref{ssec:combinedlimit}).  These distributions are normalized to unity to show the fine detail, rather than being normalized as true joint probability distributions (see text for more discussion on this).  These therefore show how combining the lower limit from Leo T and the upper limits from the \ac{GALFA-HI} and \ac{ELVIS} analysis provides an actual estimate of $M_c$ rather than only limits.}
\label{fig:infercombo}
\end{center}
\end{figure}

With an upper limit on the $\zre$ $M_c$ set by the above procedure for Leo T, and lower limit set by the above \ac{GALFA-HI} observation comparison, the joint probability of the two together provides an estimate of $M_c$. Precisely this exercise is demonstrated in Figure \ref{fig:infercombo}.  We show this for both the ``Not-Now'' (top, red) and the ``Never'' (bottom, orange) scenarios.  While the latter scenario has a notably wider distribution due to the more conservative assumptions built into it, both provide a constraint on the halo mass of reionization around $3 \times 10^8 \; M_\odot$, although potentially from $10^8$ to $10^9 \; M_\odot$. But unlike other estimates for  $M_c$, this estimate depends on no assumptions about the effects of reionization on star formation - rather this estimate derives (solely) from observations of the \HI{} content of the $z=0$ \ac{LG} galaxies.

\begin{figure}[htb]
\begin{center}
\includegraphics[width=1\columnwidth]{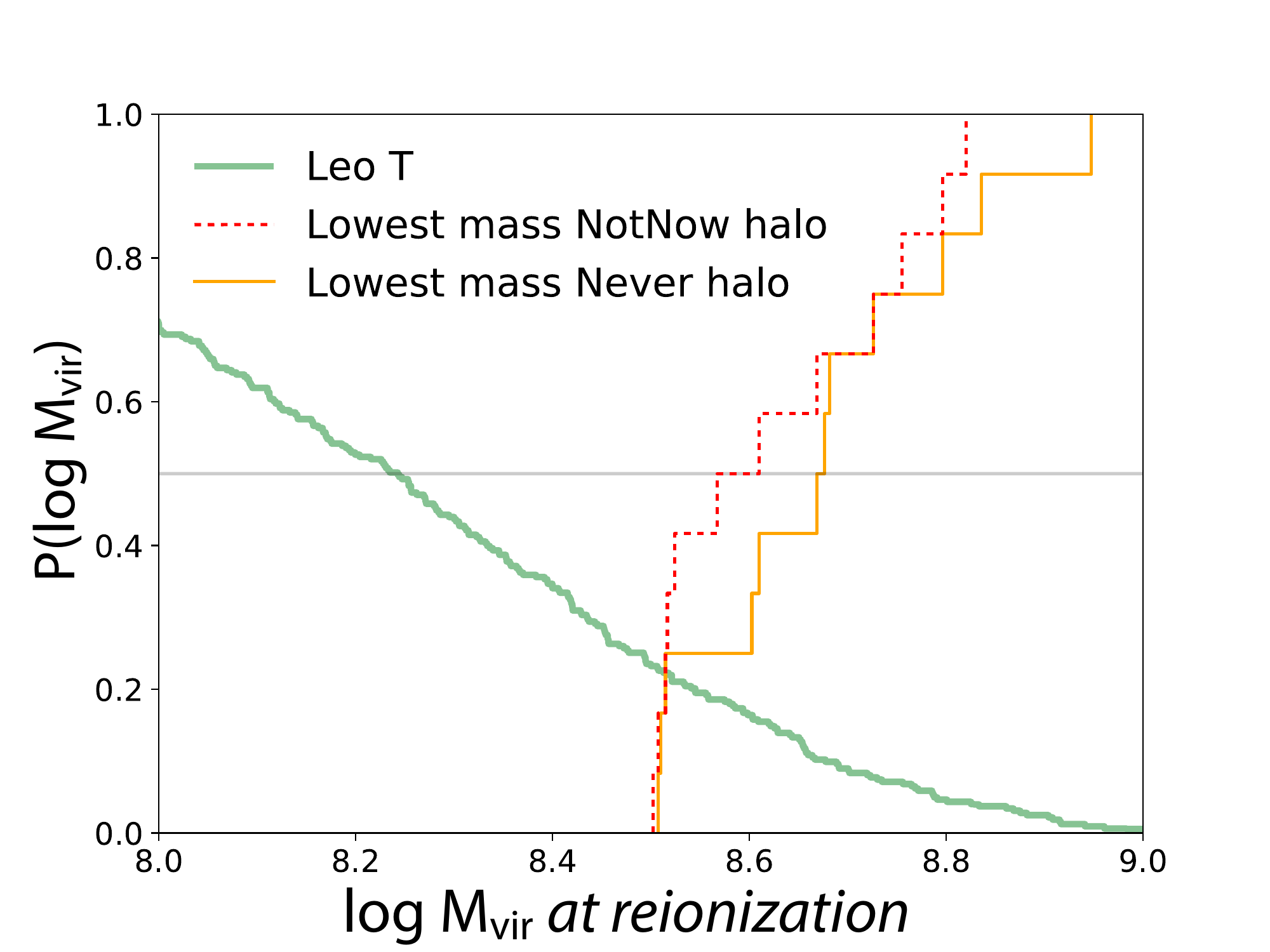}
\caption{Probability distributions (over the ELVIS simulations) of virial mass at reionization for halos that end up at $z=0$ as the lowest-mass \HI{}-bearing galaxies in the simulation, for the ``NotNow'' and ``Never'' scenarios (red and orange, respectively).  Both are computed assuming $M_c$ is the median from the joint distributions of Figure \ref{fig:infercombo}.  These halos can be interpreted as Leo T analogs under the assumption that Leo T truly is the lowest-mass gas-bearing galaxy in the Local Group.  Also shown as the green line is the at reionization virial mass distribution for Leo T (identical to the green lines in Figure \ref{fig:infercombo}).  It is clear that these distributions only weakly overlap, highlighting the tension between Leo T's existence and the \ac{GALFA-HI} observations (see text in \S \ref{ssec:combinedlimit} for more details).}
\label{fig:lowestmassleots}
\end{center}
\end{figure}

We also note a possibly surprising feature of Figure \ref{fig:infercombo}.  In both scenarios, the relative overlap of the probability distributions from the Leo T upper bound and the  \ac{GALFA-HI} lower bound is quite limited.  That is, the \emph{absolute} probability of both being correct is relatively low.  While the joint probability shown in Figure \ref{fig:infercombo} has been normalized to unity for clarity, the absolute values are quite low.  This implies that, fundamentally, there is a tension between the \ac{GALFA-HI} observations and the very existence of  Leo T's \HI{} (particularly for the ``never'' scenario). This tension persists even if we compare Leo T to only the \emph{lowest} mass \HI{}-bearing galaxy surviving through the models described in \label{ssec:galfalimit} - this experiment is illustrated inFigure \ref{fig:lowestmassleots}.  While these model Leo T analogs are on average lower mass than the Never/NotNow distributions shown in Figure \ref{fig:infercombo}, they are still in tension with the Leo T distribution.  While this could be due to inadequacies in the assumptions baked into the models used to infer $M_c$, it may also imply that Leo T truly is at the edge of the stochastic regime suggested by \citet{fitts17}, and therefore is at a mass where reionization can have a major impact on galaxy formation.

\section{Conclusions}
\label{sec:conc}

In this paper, we:

\begin{itemize}
  \item Described a process that takes dark matter simulations analogous to  the local group and, using scaling laws and well-characterized sensitivity functions, generates mock HI galaxy catalogs.
  \item Applied this transformation to the \ac{ELVIS} suite of simulations removing dwarf galaxies likely to have been stripped by interaction with the host. We find more HI observable gas-rich dwarf galaxies in all simulations than we see in \ac{GALFA-HI} CCC. At face value this suggests a significant tension between the observations and the simulations.
  \item This non-detection of gas-rich dwarf galaxies expected by simulations can be interpreted as a lower limit of the mass-scale of reionization ($\sim 10^{8.5} M_\odot$), \emph{independent} of any assumption about the impact of reionization on star formation.
  \item Combining this limit with the limit inferred from the very existence of the gas-rich dwarf galaxy Leo T, we infer the mass-scale at which reionization significantly affects dwarf galaxy formation.  This scale is consistent with those inferred by theoretical estimates.
\end{itemize}

The final point also serves to explain the relative paucity of new discoveries of gas-rich \ac{LG} dwarf galaxies despite extensive \HI{} surveys.  While here it is cast as a \emph{limit} on reionization, reversing the conclusion yields the result that if reionization becomes significant at roughly the Leo T mass scale, reionization has suppressed these galaxies' gas content.  This explains their absence in the \HI{} surveys.

While these results are at the limits of what is achievable with current \HI{} and optical surveys, new prospects are on the horizon. \ac{FAST} with its higher resolution, larger field of regard, and larger multiplexing factor, will allow us to conduct an order of magnitude more powerful study and further constrain the history of reionization. It will also provide a way to test the predictions of the simple model posed here, as new gas-bearing dwarf galaxy detections (or lack thereof) will provide cross-checks on the results presented here. \ac{FAST} is beginning its science surveys next year, and should support experiments similar to the \ac{GALFA-HI} analysis presented here.  Further afield, combined with focused observations or large, deep optical surveys like the \ac{LSST}, the techniques laid out in this paper will provide an excellent opportunity to improve the constraints on dwarf galaxy formation and reionization.

\acknowledgments
The authors thank Mike Boylan-Kolchin, James Bullock, Andrew Hearin, Justin Read, and Kirill Tchernyshyov for helpful discussions regarding this work.  The authors also thank the anonymous referee for suggestions that improved this work.  This research made use of Astropy, a community-developed core Python package for Astronomy \citep{astropy}.  It also used  the open-source software tools Numpy, Scipy, Matplotlib, and IPython \citep{numpyscipy, matplotlib, ipython}.
This research has made use of NASA's Astrophysics Data System.

Some support for EJT was provided by a Giacconi Fellowship.

\bibliographystyle{yahapj}
\bibliography{biblio}

\end{document}